\pdfoutput=1

\documentclass{sig-alternate-05-2015}

\usepackage{times}
\usepackage{graphicx} 
\usepackage{subfigure} 

\usepackage[numbers]{natbib}
\usepackage{multirow}
\usepackage{algorithm}
\usepackage{algorithmic}

\usepackage[normalem]{ulem}

\DeclareFixedFont{\ttb}{T1}{txtt}{bx}{n}{9} 
\DeclareFixedFont{\ttm}{T1}{txtt}{m}{n}{9}  

\usepackage{color}
\definecolor{deepblue}{rgb}{0,0,0.5}
\definecolor{deepred}{rgb}{0.6,0,0}
\definecolor{deepgreen}{rgb}{0,0.5,0}
\usepackage{listings}

\newcommand\pythonstyle{\lstset{
language=Python,
basicstyle=\ttm,
otherkeywords={self},             
keywordstyle=\ttb\color{deepblue},
emph={MyClass,__init__},          
emphstyle=\ttb\color{deepred},    
stringstyle=\color{deepgreen},
showstringspaces=false,
escapechar=@
}}

\lstnewenvironment{python}[1][]
{
\pythonstyle
\lstset{#1}
}
{}

\newcommand{\listemph}[1]{{\ttm \large \color{red} \bf #1}}
\newcommand{\listrepl}[1]{{\ttm \small \color{blue} \fbox{\bf #1}}}

\renewcommand{\t}[1]{{\small \textsf{#1}}}

\newcommand{\figlabel}[1]{\label{fig:#1}}
\newcommand{\figref}[1]{Figure~\ref{fig:#1}}

\newcommand{\tablabel}[1]{\label{tab:#1}}
\newcommand{\tabref}[1]{Table~\ref{tab:#1}}

\newcommand{\fixalgo}{\textsc{SynFix}}

\begin{document} 

\setcopyright{acmcopyright}
\title{Automated Correction for Syntax Errors in Programming Assignments using Recurrent Neural Networks}
\numberofauthors{2}

\author{
\alignauthor
Sahil Bhatia\\
       \affaddr{Netaji Subhas Institute of Technology}\\
       \affaddr{Delhi, India}\\
       \email{sahilbhatia.nsit@gmail.com}
\alignauthor
Rishabh Singh\\
       \affaddr{Microsoft Research}\\
       \affaddr{Redmond, WA, USA}\\
       \email{risin@microsoft.com}
}

\maketitle

\begin{abstract} 
We present a method for automatically generating repair feedback for syntax errors for introductory programming problems. Syntax errors constitute one of the largest classes of errors (34\%) in our dataset of student submissions obtained from a MOOC course on edX. The previous techniques for generating automated feedback on programming assignments have focused on functional correctness and style considerations of student programs. These techniques analyze the program AST of the program and then perform some dynamic and symbolic analyses to compute repair feedback. Unfortunately, it is not possible to generate ASTs for student programs with syntax errors and therefore the previous feedback techniques are not applicable in repairing syntax errors.

We present a technique for providing feedback on syntax errors that uses Recurrent neural networks (RNNs) to model syntactically valid token sequences. Our approach is inspired from the recent work on learning language models from Big Code (large code corpus). For a given programming assignment, we first learn an RNN to model all valid token sequences using the set of syntactically correct student submissions. Then, for a student submission with syntax errors, we query the learnt RNN model with the prefix token sequence to predict token sequences that can fix the error by either replacing or inserting the predicted token sequence at the error location. We evaluate our technique on over $14,000$ student submissions with syntax errors.
Our technique can completely repair 31.69\% (4501/14203) of submissions with syntax errors and in addition partially correct 6.39\% (908/14203) of the submissions.
\end{abstract} 

\section{Introduction}

With the ever-increasing role of computing, there has been a tremendous growth in interest in learning programming and computing skills. The computer science enrollments in universities has been growing steadily and it is becoming more and more challenging to meet this increasing demand. Recently, several online education initiatives such as edX, Coursera, and Udacity have started providing Massive Open Online Courses (MOOCs) to tackle this challenge of providing quality education at scale that is easily accessible to students worldwide. While there are several benefits of MOOCs -- access to quality course material and instruction, cheaper than traditional university courses, learning at one's own pace etc., there are also several drawbacks. One important drawback is that students enrolled in MOOCs do not typically get quality feedback for assignments compared to the feedback provided in traditional classroom settings since it is prohibitively expensive to hire enough instructors and teaching assistants to provide individual feedback to thousands of students. In this paper, we address the problem of providing automated feedback on \emph{syntax errors} in programming assignments using machine learning techniques.

The problem of providing feedback on programming assignments at scale has seen a lot of interest lately. These approaches can be categorized into two broad categories -- peer-grading~\cite{peergrading} and automated grading techniques~\cite{autoprof,codewebs}. In the peer-grading approach, students rate and provide feedback on other student submissions based on a grading rubric. Some MOOCs have made this step mandatory for students before they get feedback for their own assignments. While peer-grading has been shown to be effective for student learning, it also presents several challenges. First, it takes a long time (sometimes days) to get any useful feedback, and second, there is a potential for inaccuracies in feedback especially when students providing feedback themselves are struggling in learning the material.

The second approach of automated feedback generation aims to automatically provide feedback on student submissions. Most recent approaches for automated grading have focused on providing feedback on the functional correctness and style considerations of student programs. AutoProf~\cite{autoprof} is a system for providing automated feedback on functional correctness of buggy student solutions. It uses constraint-based synthesis techniques to find minimum number of changes to an incorrect student submission such that it becomes functionally equivalent to a reference teacher implementation. The Codewebs system~\cite{codewebs} is a search engine for coding assignments that allows for querying massive dataset of student submissions using "code phrases", which are subgraphs of AST in the form of subtrees, subforests, and contexts. A teacher provides feedback on a handful of submissions, which is then propagated to provide feedback on thousands of submissions by querying the dataset using code phrases. 

While providing feedback on functional and stylistic elements of student submissions is important, a significant fraction of submissions (more than 34\% in our dataset) comprise of syntax errors and providing feedback on syntactic errors has largely been unexplored. Many of the techniques described previously for automated grading can not provide feedback on syntactic errors since they inherently depend on analyzing the AST of the student submission, which is unfortunately not available for programs with syntax errors. Although compilers have improved a lot in finding the error location and in describing the syntax errors using better error messages, they can not provide feedback on how to fix these errors in general since they are developed for general-purpose scenarios.

In this paper, we present a technique to automatically provide feedback on student programs with syntax errors leveraging the large dataset of correct student submissions. Our hypothesis is that even though there are thousands of student submissions, the diversity of solution strategies for a given problem is relatively small and the fixes to syntactic errors can be learnt from correct submissions. For a given programming problem, we use the set of (possibly functionally incorrect) student submissions without syntax errors to learn a sequence model of tokens, which is then used to hypothesize possible fixes to syntax errors in a student solution. Our system incorporates the suggested changes to the incorrect program and if the modified program passes the compiler syntax check, it provides those changes as possible fixes to the syntax error. We use a Recursive Neural Network (RNN)~\cite{rnn} to learn the token sequence model that can learn large contextual dependencies between tokens.

Our approach is inspired from the recent pioneering work on learning probabilistic models of source code from a large repository of code for many different applications~\cite{naturalness,nguyen,codemining,namesuggestion,codeconventions,idioms,bigcode}. Hindle et al.~\cite{naturalness} learn an n-gram language model to capture the repetitiveness present in a code corpora and show that n-gram models are effective at capturing the local regularities. They used this model for suggesting next tokens that was already quite effective as compared to the type-based state-of-the-art IDE suggestions. Nguyen et al.~\cite{nguyen} enhanced this model for code auto-completion to also include semantic knowledge about the tokens (such as types) and the scope and dependencies amongst the tokens to consider global relationships amongst them. The \textsc{Naturalize} framework~\cite{codeconventions} learns an n-gram language model for learning coding conventions and suggesting changes to increase the stylistic consistency of the code. More recently, some other probabilistic models such as conditional random fields and log bilinear context models have been presented for suggesting names for variables, methods, and classes~\cite{bigcode,namesuggestion}. We also learn a language model to encode the set of valid token sequences, but our approach differs from these approaches in four key ways: i) our application of using a language model learnt from syntactically correct programs to fix syntax errors is novel and different from previous applications, ii) since we cannot obtain the Abstract Syntax Tree (AST) of these programs with syntax errors many of these techniques that use AST information for learning the language model are not applicable, iii) we learn recursive neural networks (RNN) that can capture more complex dependencies between tokens than n-gram or logbilinear neural networks, and finally iv) instead of learning one language model for the complete code corpus, we learn individual RNN models for different programming assignments so that we can generate individualized repair feedback for different problems.

We evaluate the effectiveness of our technique on student solutions for 5 programming problems taken from the Introduction to Programming class (6.00x) offered on the edX platform. Our technique can suggest tokens for completely fixing the syntax errors for $31.69\% (4501/14203)$ of the submissions with syntax errors. Moreover, for an additional $6.39\% (908/14203)$ programs, our technique can suggest fixes that correct the first syntax error in the program but doesn't fully correct the program because of the presence of multiple syntax errors. 

This paper makes the following key contributions:

\begin{itemize}
\item We formalize the problem of finding fixes for syntax errors in student submissions as a token sequence learning problem using the recurrent neural networks (RNN).
\item We present the $\fixalgo$ algorithm to use the predicted token sequences for finding repairs to syntax errors that performs different code transformations including insertion and replacement of predicted sequences in a ranked order.
\item We evaluate the effectiveness of our system on more than $14,000$ student submissions from an online introductory programming class. Our system can completely correct the syntax errors in $31.69\%$ of the submissions and partially correct the errors in an additional $6.39\%$ of the submissions.
\end{itemize}
 
\section{Motivating Examples}
We now present a few examples of the different types of syntax errors we encounter in student submissions from our dataset and the repair corrections our system is able to generate using the token sequence model learnt from the syntactically-correct student submissions. The example corrections are shown in \figref{motivatingexamples} for the student submissions for the \t{recPower} problem taken from the Introduction to Programming MOOC (6.00x) on edX. The \t{recPower} problem asks students to write a recursive Python program to compute the value of $\t{base}^\t{exp}$ given a real value \t{base} and an integer value \t{exp} as inputs.

Our syntax correction algorithm considers two types of parsing errors in Python programs: i) Syntax errors, and ii) Indentation errors. It uses the offset information provided by the Python compiler to locate the potential locations for syntax errors, and then uses the program statements from the beginning of the function to the error location as the prefix token sequence for performing the prediction. However, there are many cases such as the ones shown in \figref{motivatingexamples}(c) where the compiler is not able to accurately find the exact offset location for the syntax error. In such cases, our algorithm ignores the tokens present in the error line and considers the prefix ending at the previous line. Using the prefix token sequence, the algorithm uses a neural network to perform the prediction of next $k$ tokens that are most likely to follow the prefix sequence, which are then either inserted at the error location or are used to replace the original token sequence at the error location.

A sample of syntax errors and the fixes generated by our algorithm (emphasized in boldface red font) based on inserting the predicted tokens from the offset location is shown in \figref{motivatingexamples}(a). For correcting syntax errors in this class, our algorithm first queries the learnt language model to predict the next token sequence using the prefix token sequence ending at either the offset (error) location or one token before the offset location (Offset-1). It then tries inserting the tokens from the predicted sequence in increasing order of length at the corresponding offset location until the syntax error in the line is fixed. The kinds of errors in this class typically include inserting unbalanced parenthesis, completing partial expressions (such as \t{exp-} to \t{exp-1}), adding syntactic tokens such as \t{:} after \t{if} and \t{else} expressions, etc. 

Some example syntax errors that require replacing the original tokens in the incorrect program with the predicted tokens are shown in \figref{motivatingexamples}(b). These errors typically include replacing an incorrect operator with another operator (such as replacing \t{=} with \t{*}, \t{=} in comparisons with \t{==}), deleting additional mismatched parenthesis etc. Our algorithm performs a similar technique for generating prefix token sequences as in the case of previous class of syntax errors that require token insertion. The only difference is that instead of inserting the tokens from the predicted token sequence, it replaces the original tokens with the predicted tokens.

There are several cases in which the Python compiler isn't able to accurately locate the error location offset. Some examples of these cases are shown in \figref{motivatingexamples}(c) that include wrong spelling of keywords (\t{retrun} instead of \t{return}, \t{f} instead of \t{if}), wrong expression for the return statement etc. For fixing such syntax errors, our algorithm generates the prefix token sequence that ends at the last token of the line previous to the error line and ignores all the tokens occurring in the error line. It then queries the model to predict a token sequence that ends at a new line, and then replaces the error line completely with the predicted token sequence.

Finally, a sample of indentation errors is shown in \figref{motivatingexamples}(d). These errors typically involve mistyped operators, using the wrong indentation after a function definition, conditional, and loop expressions etc. Our algorithm tries the same strategy as described previously for the class of syntax errors including inserting or replacing the tokens at the offset location.

An interesting point to note here is that currently our system predicts token sequences for fixing the syntax errors in the code that may or may not correspond to the correct \emph{semantic} fix, i.e. the suggested fix would pass the parser check but may not compute the desired result (or may even throw a runtime exception). For example, in some cases such as the incorrect expression \t{recurPower(base,exp-=1)}, the top token sequence prediction results in the expression \t{recurPower(base,exp-11)}, which is a syntactically correct expression but does not result in computing the desired result of computing $\t{base}^\t{exp}$. Even for such cases, the generated fix can still provide some hints to the students about the correct usage of expressions for the corresponding program contexts. However, for many of the cases, the suggested repair for syntax correction also happens to correspond to the correct semantic fix as shown in \figref{motivatingexamples}.

\begin{figure*}
\begin{tabular}{|c | c|}
\hline
\multicolumn{2}{|c|}{\bf (a) SyntaxError - Insert Token}\\
\hline
\begin{minipage}{0.5\linewidth}
{\scriptsize
\begin{python}
def recPower(base, exp):
    if exp <= 0:
        return 1
    return base * recPower(base, exp - 1 @\listemph{)}@    
\end{python}
}
\end{minipage}
&
\begin{minipage}{0.5\linewidth}
{\scriptsize
\begin{python}
def recPower(base, exp):
    if exp <= 0:
        return 1
    return base * recPower(base, exp-@\listemph{1}@)    
\end{python}
}
\end{minipage}

\\ \hline

\begin{minipage}{0.5\linewidth}
{\scriptsize
\begin{python}
def recPower(base, exp):
    if exp == 1:
        return base
    return base * (recPower(base, (exp - 1))@\listemph{)}@    
\end{python}
}
\end{minipage}

&
\begin{minipage}{0.5\linewidth}
{\scriptsize
\begin{python}
def recPower(base, exp):
    if exp > 1:
        return base * recurPower(base, exp-1)
    else@\listemph{:}@
        return 1
\end{python}
}
\end{minipage}
\\\hline

\multicolumn{2}{|c|}{\bf (b) SyntaxError - Replace Token}\\
\hline
\begin{minipage}{0.5\linewidth}
{\scriptsize
\begin{python}
def recurPower(base, exp):
    if exp == 0:
        return 1
    return base @\listrepl{=} \listemph{* } @recurPower(base,exp-1)
\end{python}
}
\end{minipage}

&

\begin{minipage}{0.5\linewidth}
{\scriptsize
\begin{python}
def recurPower(base, exp):
    total = base
    if(exp==0):
        return total
    else:
        total*=base
        return total+recurPower(base,exp-@\listrepl{=}\listemph{1}@1)
\end{python}
}
\end{minipage}

\\ \hline

\begin{minipage}{0.5\linewidth}
{\scriptsize
\begin{python}
def recurPower(base, exp):
    if exp@\listrepl{=} \listemph{==}@0:
      return 1;
    else:
      return base*recurPower(base,exp-1)
\end{python}
}
\end{minipage}

&

\begin{minipage}{0.5\linewidth}
{\scriptsize
\begin{python}
def recurPower(base, exp):
    if exp == 0:
        return 0
    elif exp == 1:
        return base
    else:
        return base*recurPower(base,exp-1)@\listrepl{)}@
\end{python}
}
\end{minipage}
\\ \hline

\multicolumn{2}{|c|}{\bf (c) SyntaxError - Previous Line Insert}\\
\hline

\begin{minipage}{0.5\linewidth}
{\scriptsize
\begin{python}
def recurPower(base, exp):
    @\listrepl{f exp == 1:}@ 
    @\listemph{if exp == 1:}@
        return base
    return base * recurPower(base, (exp - 1))
\end{python}
}
\end{minipage}

&

\begin{minipage}{0.5\linewidth}
{\scriptsize
\begin{python}
def recurPower(base, exp):
    if exp == 1:
        return base
    else:
        @\listrepl{retrun base * recurPower(base, exp - 1)}@
        @\listemph{return base * recurPower(base, exp - 1)}@
\end{python}
}
\end{minipage}

\\ \hline

\begin{minipage}{0.5\linewidth}
{\scriptsize
\begin{python}
def recurPower(base, exp):
    if exp == 0:
        @\listrepl{return = exp + 1}@
        @\listemph{return base}@
    else:
        return (base*recurPower(base,exp-1))
\end{python}
}
\end{minipage}

&

\begin{minipage}{0.5\linewidth}
{\scriptsize
\begin{python}
def recurPower(base, exp):
    if exp == 0:
        return 1
    if exp == 1:
        return base
    if exp > 1:
        @\listrepl{return exp -= 1}@
        @\listemph{return base * recurPower(base, exp-1)}@
    else:
        return recurPower(base,exp-1)
\end{python}
}
\end{minipage}

\\ \hline

\multicolumn{2}{|c|}{\bf (d) Indentation Error - Insert Token}\\
\hline
\begin{minipage}{0.5\linewidth}
{\scriptsize
\begin{python}
def recurPower(base, exp):
    if exp == 0:
    @\listrepl{return 1}@
    @\listemph{\;\;\;\;\;return 1}@
    return base * recurPower(base,exp-1)
\end{python}
}
\end{minipage}

&

\begin{minipage}{0.5\linewidth}
{\scriptsize
\begin{python}
def recurPower(base, exp):
    x = base
    while(exp > 0):
        x *= base
	@\listrepl{-= 1}@
	@\listemph{exp -= 1}@
    return base
\end{python}
}
\end{minipage}

\\ \hline
\end{tabular}
\caption{Some examples of the fixes suggested by our system for the \t{recurPower} student submissions with syntax errors taken from edX. The suggestions are emphasized in red using larger font, whereas the the program expressions/statements that are removed are emphasized in blue with a frame box.}
\figlabel{motivatingexamples}
\end{figure*}
\section{Approach}
An overview of the workflow of our system is shown in \figref{overview}. For a given programming problem, we first use the set of all syntactically correct student submissions to train a neural network in the training phase for learning a token sequence model for all valid token sequences that is specific to the problem. We then use the $\fixalgo$ algorithm to find small corrections to a student submission with syntax errors using the token sequences predicted from the learnt model. These corrections are then used for providing feedback in terms of potential fixes to the syntax errors. We now describe the two key phases in our workflow: i) the training phase, and ii) the $\fixalgo$ algorithm.

\begin{figure}
\centering
\includegraphics[scale=0.85]{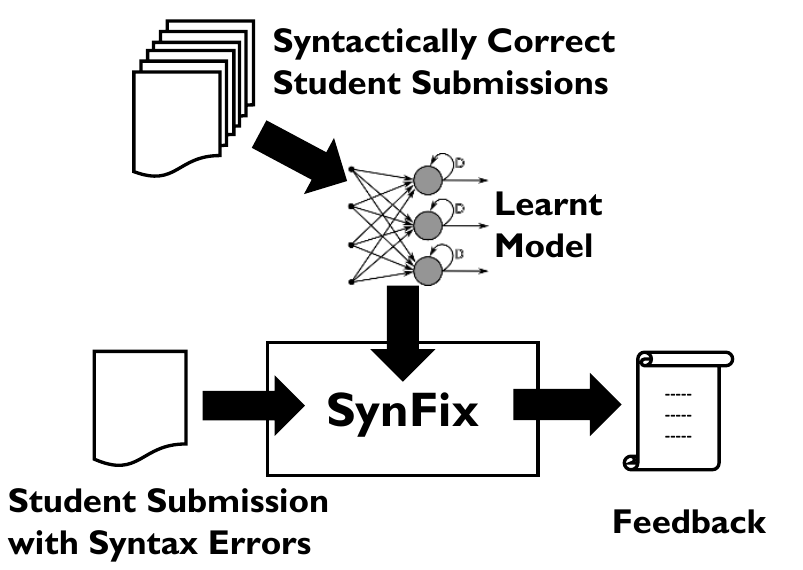}
\caption{An overview of the workflow of our system.}
\figlabel{overview}
\end{figure}

\begin{figure*}
\begin{tabular} {c c}
\begin{minipage}{0.6\linewidth}
\includegraphics[scale=0.3]{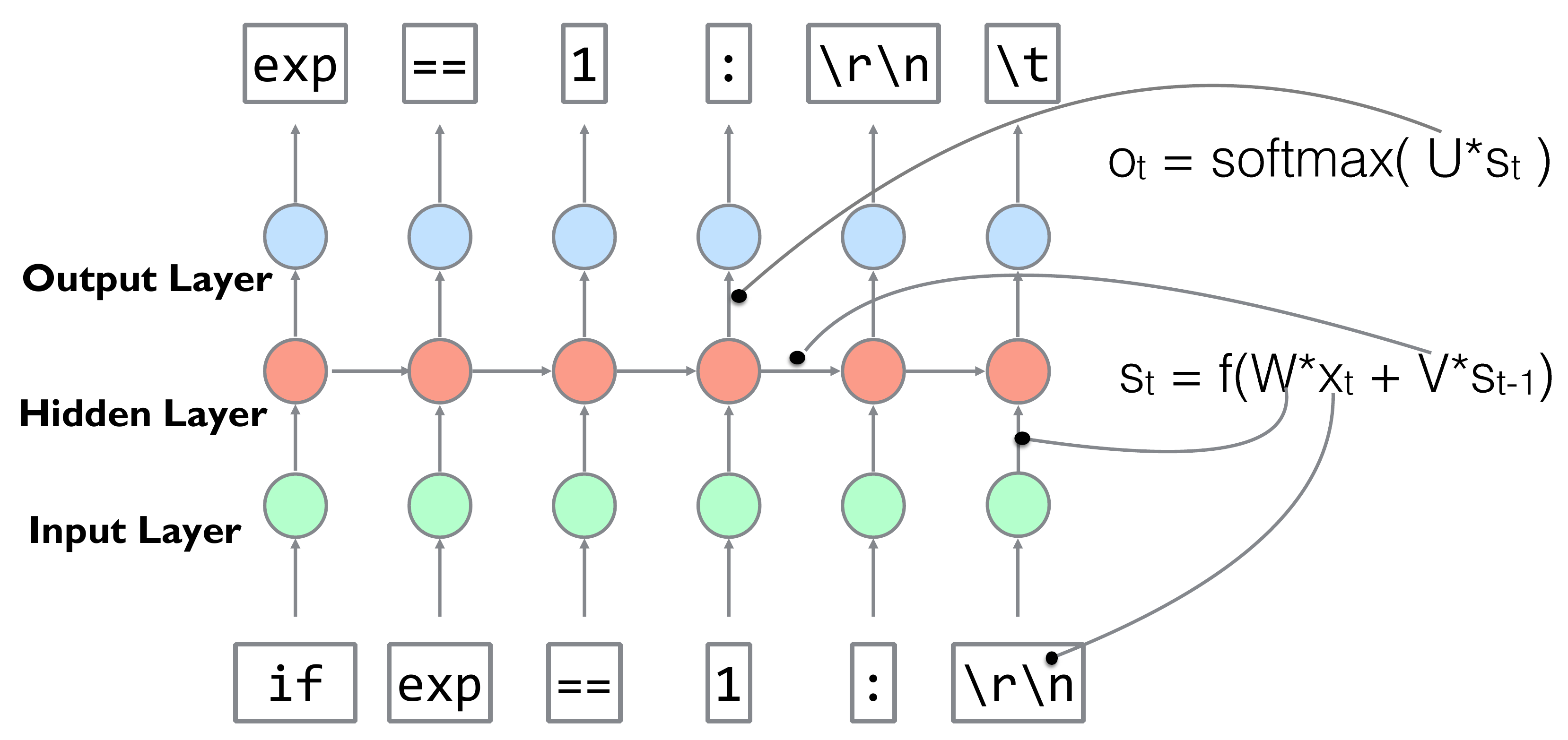}
\end{minipage}
&
\begin{minipage}{0.4\linewidth}
\includegraphics[scale=0.3]{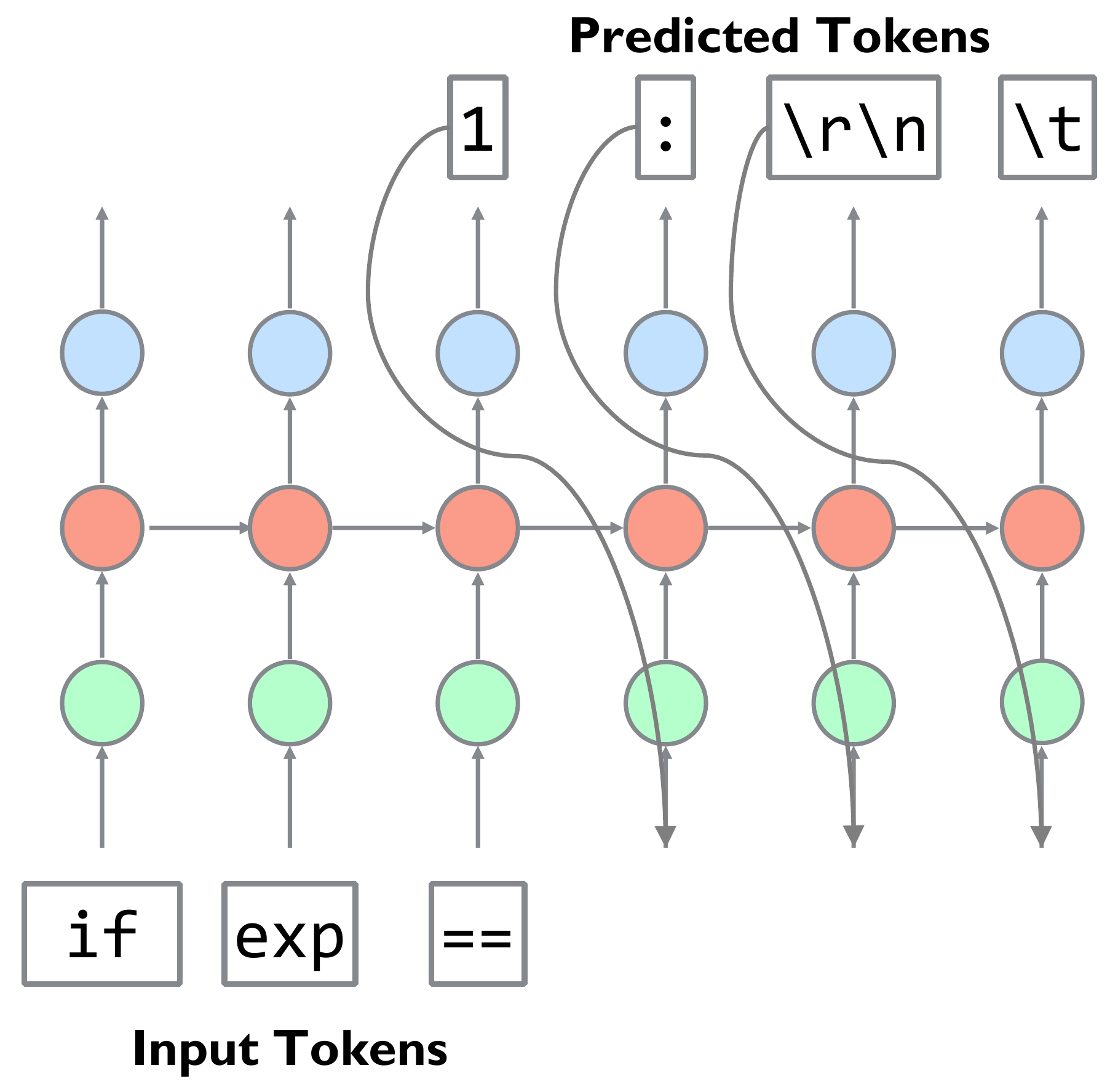}
\end{minipage}
\\
(a) Training Phase
&
(b) Prediction Phase
\end{tabular}
\caption{The modeling of our syntax repair problem using an RNN with 1 hidden layer. (a) We provide input and output token sequences in the training phase to learn the weight matrices. (b) In the prediction phase, we provide a token sequence to the input layer of the RNN and generate the output token sequences using the learnt model.}
\figlabel{rnnexample}
\end{figure*}

\subsection{Neural Network Model}
The simplest class of neural networks~\cite{neuralnetworks} (also called convolutional networks) are feedforward neural networks and were the first type of artificial neural network devised. These networks accept a fixed-sized vector as input (e.g. a bag of words model of a piece of text) and produce a fixed-size vector as output (e.g. the sentiment label for the text). In these networks, the information moves in only one forward direction from the input nodes to the hidden layers to the output layer. The feedforward networks have been found to be quite successful for a variety of classification tasks including sentiment analysis, image recognition, document relevance etc. However, there are two big limitations of these networks: 1) they only accept fixed-size input vectors and 2) they can perform only a fixed number of computational steps (defined by the fixed number of hidden layers). 

To overcome these limitations, another class of neural networks called RNN (Recurrent Neural Network) have been devised that can operate over sequences of input and output vectors as opposed to fixed-length vectors. Moreover, in addition to the feedforward structure of the network, the output of a hidden layer is connected to its own input (cyclic paths) thus generating a feedback in the network. This feedback property of RNN gives them memory to retain information from previous steps and then use it for processing the current and future states. These additional capabilities make RNNs a very powerful computational model and can theoretically represent long context patterns. Although the RNNs are much more expressive than n-gram and feed forward networks, the conventional wisdom has been that RNNs are more difficult to train. But with some recent algorithmic and computational advances, they have been shown to be efficiently learnable and have recently been used successfully for many tasks such as machine translation, video classification by frames, speech recognition etc. In this section, we describe how we model our problem of learning token sequences from syntactically correct student programs and then predicting token sequences for repairing incorrect programs using RNNs.

\begin{figure}
\centering
\includegraphics[scale=0.42]{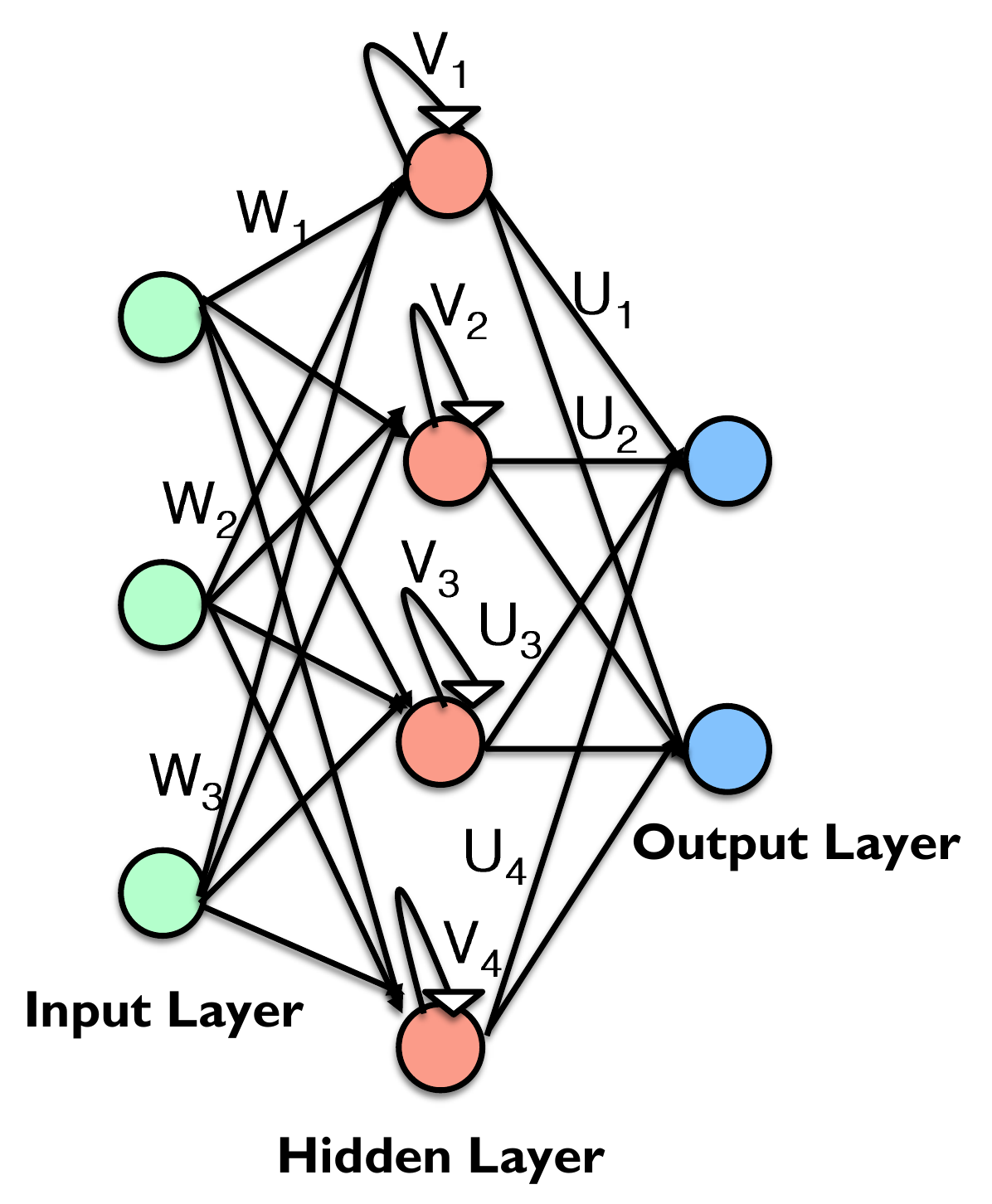}
\caption{A simple RNN with a single hidden layer.}
\figlabel{simplernn}
\end{figure}

We first describe a brief overview of the computational model of a simple RNN with a single hidden layer. Consider an input sequence of length $L$ and an RNN with $I$ number of inputs, a single hidden layer with $H$ number of hidden units, and $k$ output units. Let $x_t \in R^I$ denote the input at time step $t$ (encoded as a vector), $s_t \in R^H$ denote the hidden state at time step t, $W\in R^{H \times I}$ denote the weight matrix corresponding to the weights on connections from input layer to hidden layer, $V \in R^{H \times H}$ be the weight matrix from hidden to hidden layer (recursive), and $U \in R^{I \times H}$ be the weight matrix from hidden to the output layer. A simple RNN architecture with $I=3$ number of inputs, $H=4$ number of hidden units in a single hidden layer, and $k=2$ output units is shown in \figref{simplernn}. The computation model of the RNN can be defined using the following equations:
\begin{align*}
s_t = f(W*x_t + V*s_{t-1}) \\
o_t = \t{softmax}(U*s_t)
\end{align*}

The hidden state vector $s_t$ at time step $t$ is computed by applying an activation function $f$ (e.g. \t{tanh} or \t{sigmoid}) to a weighted sum of the input vector $x_t$ and the previous hidden state vector $s_{t-1}$. The output vector $o_t$ is computed by applying the \t{softmax} function to the weighted state vector value $s_t$.

The artificial neurons are analogous to the neurons in human body which continuously receive electrochemical signal through their dendrites and when the sum of these signal surpass a certain threshold they send(fire) the electrochemcial signals through their axons. The hidden units and output units use a similar activation strategy to determine the state of the units during a particular time stamp. The hidden units take the weighted sum as input and map it to a value in the set (-1,1) using the sigmoid function to model non-linear activation relationships. The activation of a unit $h$ in the hidden layer and output unit $k$ at time step $t$ is given by:

\begin{align*}
a_t^h = \displaystyle\sum_{i=1}^{I} W[i,n] * x^{i}_{t} + \sum_{h'=1}^{H} V[h',h] * s^{h'}_{t-1}\\
a_t^k = \sum_{h=1}^{H} U[n,k] * f(a_t^h)
\end{align*}

During the training phase, RNN uses backpropagation through time(BPTT)~\cite{bptt} to calculate the gradient and adjust the weights. BPTT is an extension of the backpropagation algorithm that takes into account the recursive nature of the hidden layers from one time step to the next. The loss function depends not only on the direct influence of the hidden layer but also on the values from hidden layer during the next time step. The loss function which is minimized during the training is the cross entropy error between the training output label and the predicted output label.

There are two common ways to feed each word of the sequence to the input layer of the RNN: 1) words are represented as one hot vector which is multiplied by the weight matrix and used for the forward pass, and 2) words are mapped to high dimensional vector and an embedding matrix is used to perform lookups. While training the network for learning sequence models, the target sequence is the input sequence shifted left by 1 since the model is trained to minimize the negative log likelihood of the predicted token and the next actual token in the sequence.

We now describe how we model our syntax repair problem for a given programming assignment using an RNN. We first use the syntactically correct student submissions to obtain the set of all valid token sequences. We then use a threshold frequency value to decide whether to relabel a token to a general \t{IDENT} token for handling rarely used tokens (such as infrequent variable/method names). A token is encoded into a fixed-length hot vector such that it contains $1$ for the index corresponding to the token index in the vocabulary and $0$ in all other places. The size of the hot vector is equal to the size of the training vocabulary. 

In the training phase, we provide the token sequences to the input layer of the RNN and the input token sequence shifted left by 1 as the target token sequence to the output layer as shown in \figref{rnnexample}(a). The figure also shows the equations to compute the output probabilities for output tokens and the weights associated with connections from input to hidden layer, hidden to hidden layer, and hidden to output layer. After learning the network from the set of syntactically correct token sequences, we use the model to predict next token sequences given a prefix of the token sequence to the input layer as shown in \figref{rnnexample}(b). The first output token is predicted at the output layer using the input token sequence. For predicting the next output token, the predicted token is used as the next input token in the input layer as shown in the figure.

{\bf Long Short Term Memory networks (LSTM)}: LSTMs~\cite{lstm} are a special kind of RNN that are capable of learning long-term dependencies and have been shown to outperform general RNNs for a variety of tasks. In theory, RNNs are capable of handling any form of long-term dependencies on the past information because of the recursive connections. But, in practice, RNNs only perform well for cases where the gap between the required context information and the place where it's needed is small. As the gap becomes larger, it becomes more difficult for RNNs to learn to connect the desired information. LSTMs are explicitly designed to avoid this long-term dependency issue with the RNNs. Instead of regular network units, the LSTMs contain LSTM blocks that intuitively determine whether the input is significant enough to remember, when it should forget the value, and when the value should be used for other layers. In the evaluation section, we also use different LSTM models to learn token sequences and compare their performance with the RNN models.

\subsection{The SYNFIX Algorithm}
The $\fixalgo$ algorithm, shown in Algorithm.~\ref{fixalgo}, takes as input a program $P$ (with syntax errors) and a token sequence model $\mathcal{M}$, and returns either a fixed program $P'$ (if possible) or $\phi$ denoting that the program cannot be fixed. The algorithm first uses a parser to obtain the type of error \t{err} and the token location where the error occurs \t{loc}, and computes a prefix of the token sequence $\widetilde{\rm T}_{prefix}$ corresponding to the token sequence starting from the beginning of the program until the error token location \t{loc}. We use the notation $a[i..j]$ to denote a subsequence of a sequence $a$ starting at index $i$ (inclusive) and ending at index $j$ (exclusive). The algorithm then queries the model $\mathcal{M}$ to predict the token sequence $\widetilde{\rm T}_k$ of a constant length $k$ that is most likely to follow the prefix token sequence.

After obtaining the token sequence $\widetilde{\rm T}_k$, the algorithm iteratively tries token sequences $\widetilde{\rm T}_k[1..i]$ of increasing lengths ($1\leq i \leq k)$ until either inserting or replacing the token sequence $\widetilde{\rm T}_k[1..i]$ at the error location results in a fixed program $P'$ with no syntax errors. If the algorithm cannot find a token sequence that can fix the syntax errors in the program $P$, the algorithm then creates another prefix $\widetilde{\rm T}_{prefix}$ of the original token sequence such that it ignores all previous tokens in the same line as that of the error token location. It then predicts another token sequence $\widetilde{\rm T}^{prev}_k$ using the model for the new token sequence prefix, and selects a subsequence $\widetilde{\rm T}^{prev}_k[1..m]$ that ends at a new line token. Finally, the algorithm checks if replacing the line containing the error location with the predicted token sequence results in no syntax errors. If yes, it returns the fixed program $P'$. Otherwise, the algorithm returns $\phi$ denoting that no fix can be found for the syntax error in $P$.

\begin{algorithm}[tb]
   \caption{$\fixalgo$}
   \label{fixalgo}
\begin{algorithmic}
   \STATE {\bfseries Input:} buggy program $P$, token sequence model $\mathcal{M}$
   \STATE $\t{(err,loc)} := \t{Parse}(P)$; $\widetilde{\rm T} := \t{Tokenize}(P)$
   \STATE $\widetilde{\rm T}_{prefix} := \widetilde{\rm T}[1..\t{loc}]$
   \STATE $\widetilde{\rm T}_k := \t{Predict}(\mathcal{M},\widetilde{\rm T}_{prefix})$
   \FOR{$i \in \t{range}(1,k)$}
   \STATE $P'_{ins} := \t{Insert}(P,\t{loc},\widetilde{\rm T}_k[1..i])$
   \STATE {\bf if} $\t{Parse}(P'_{ins}) = \phi$ {\bf return} $(P'_{ins}, \widetilde{\rm T}_k[1..i])$
   \STATE $P'_{repl} := \t{Replace}(P,\t{loc},\widetilde{\rm T}_k[1..i])$
   \STATE {\bf if} $\t{Parse}(P'_{repl}) = \phi$ {\bf return} $(P'_{repl}, \widetilde{\rm T}_k[1..i])$
   \ENDFOR
   \STATE $\widetilde{\rm T}^{prev}_{prefix} := \widetilde{\rm T}[1..\t{previousline(loc)}]$
   \STATE $\widetilde{\rm T}^{prev}_k := \t{Predict}(\mathcal{M},\widetilde{\rm T}^{prev}_{prefix})$
   \STATE $P'_{prev} := \t{ReplaceLine}(P,\t{line(loc)},\widetilde{\rm T}^{prev}_k[1..m])$
   \STATE $~~~~~~~~~~ \mbox{where~~} \widetilde{\rm T}^{prev}_k[m] = ~{\char`\\n})$
   \STATE {\bf if} $\t{Parse}(P'_{prev}) = \phi$ {\bf return} $(P'_{prev}, \widetilde{\rm T}^{prev}_k[1..m])$
\STATE {\bf return} $\phi$   
\end{algorithmic}
\end{algorithm}

{\bf Example}: Consider the Python program shown in \figref{ex1}. The Python parser returns a syntax error in line 2 with the error offset corresponding to the location of the \t{=} token. The $\fixalgo$ algorithm first constructs a prefix of the token sequence consisting of tokens from the start of the program to the error location such that $\widetilde{\rm T}_{prefix} = [\t{'def'}, \t{'recurPower'}, \t{'('}, \t{'base'}, \t{','}, \t{'exp'},\t{')'},\t{':'},\t{'}\backslash \t{r} \backslash \t{n'},$ $\t{'}\backslash \t{t'},\t{'if'}, \t{'exp'}]$. It then queries the learnt model to predict the most likely token sequence that can follow the input prefix sequence. Let us assume the value for length of predicted sequence $k$ is set to 3 and the model returns the predicted token sequence $\widetilde{\rm T}_{k}=[\t{'=='},\t{'0'},\t{':'}]$. The algorithm first tries to use the smaller prefixes of the predicted token sequence (in this case \t{'=='}) to see if the syntax error can be fixed. It first tries to insert the predicted token sequence \t{'=='} in the original program but that results in the expression \t{if exp == = 0:} that still results in an error. It then tries to replace the original token sequence with the predicted token sequence, which results in the expression \t{if exp == 0:} that passes the parser check. The algorithm then returns the corresponding feedback of replacing the token \t{'='} with the token \t{'=='} for fixing the syntax error.

Consider another incorrect Python attempt shown in \figref{ex2}, where there is a syntax error at the token \t{'exp'} in line 3 \t{retrun exp} (wrong spelling of the return keyword). The algorithm similarly constructs the prefix token sequence as $\widetilde{\rm T}_{prefix} = [\t{'def'}, \t{'recurPower'},$ $\t{'('}, \t{'base'}, \t{','}, \t{'exp'},\t{')'},\t{':'},\t{'}\backslash \t{r} \backslash \t{n'},\t{'if'},$ $\t{'exp'}, \t{'=='}, \t{'1'}, \t{'}\backslash \t{r} \backslash \t{n'}, \t{'}\backslash \t{t'},\t{'retrun'}]$. For this prefix, the algorithm is not able to either insert or replace the predicted token sequence in the original program such that the syntax error is removed. The algorithm then removes all the tokens in the prefix token sequence that occur in the error line (in this case removes the tokens $\t{'}\backslash \t{t'}$ and \t{'retrun'}), and then queries the model again to predict another token sequence with the updated prefix sequence such that the predicted sequence ends in a newline token. In this case, the algorithm predicts the token sequence corresponding to the statement \t{return base} that fixes the original syntax error.

\begin{figure}
{\scriptsize
\begin{python}
def recurPower(base, exp):
    if exp = 0:
      return 1;
    else:
      return base*recurPower(base,exp-1)
\end{python}
}
\caption{An incorrect student submission to the \t{recurPower} problem with a syntax error in line 2 (= instead of ==).}
\figlabel{ex1} 
\end{figure}

\begin{figure}
{\scriptsize
\begin{python}
def recurPower(base, exp):
    if exp == 1:
      retrun exp;
    else:
      return base*recurPower(base,exp-1)
\end{python}
}
\caption{An incorrect student submission to the \t{recurPower} problem with a syntax error in line 3 (wrong spelling of return).}
\figlabel{ex2} 
\end{figure}
\section{Evaluation}

We now present the evaluation of our system on $40,835$ Python submissions taken from the Introduction to Programming in Python course on the edX MOOC platform. The first question we investigate is whether it is possible to learn the RNN models for token sequences that can capture syntactically valid sequences. We then evaluate in how many cases our system can fix the syntax errors with the predicted sequences using different algorithmic choices in the $\fixalgo$ algorithm. Finally, we also experiment with different RNN and LSTM configurations and the vocabulary threshold value to evaluate their effect on the final result.

\subsection{Benchmarks}
Our benchmark set consists of student submissions to five programming problems \t{recurPower}, \t{iterPower}, \t{oddTuples}, \t{evalPoly}, and \t{compDeriv} taken from the edX course. The \t{recurPower} problem asks students to write a recursive function that takes as input a number \t{base} and an integer \t{exp}, and computes the value $\t{base}^{\t{exp}}$. The \t{iterPower} problem has the same functional specification as the \t{recurPower} problem but asks students to write an itervative solution instead of a recursive solution. The \t{oddTuples} problem asks students to write a function that takes as input a tuple $l$ and returns another tuple that consists of every other element of $l$. The \t{evalPoly} problem asks students to write a function to compute the value of a polynomial on a given point, where the coefficients of the polynomial are represented using a list of doubles. Finally, the \t{compDeriv} problem asks students to write a function to return a polynomial (represented as a list) that corresponds to the derivative of a given polynomial.

The number of student submissions for each problem in our evaluation set is shown in \tabref{benchmarks}. In total our evaluation set consists of $40,835$ student submissions. The number of submissions for the \t{evalPoly} and \t{compDeriv} problems are relatively lesser than the number of submissions for the other problems. This is because these problems were still in the submission stage at the time we obtained the data snapshot from the edX platform. But this also gives us a measure to evaluate how well our technique performs when we have thousands of correct attempts in the training phase as opposed to only hundreds of correct attempts. Another interesting aspect to observe from the table is the fact that a large fraction of student submissions have syntax errors ($34.78\%$). For each problem, we use the set of syntactically correct student submissions for learning the recurrent neural network and use the submissions with syntax errors as the test set to evaluate the learnt model.

\begin{table}
\begin{tabular}{|c|c|c|c|c|}
\hline
{\bf Problem} & {\bf Total} & {\bf Syntactically} & {\bf Syntax Errors}\\
& {\bf Attempts} & {\bf Correct} & {\bf (Percentage)}  \\ \hline
{recurPower} & 10247 & 8176 & 2071 (20.21\%)\\ \hline
{iterPower} & 11855 & 9194 & 2661 (22.45\%) \\ \hline
{oddTuples} & 17057 & 8233 & 8824 (51.73\%) \\ \hline
{evalPoly} & 1148 & 824 & 324 (28.22\%)\\ \hline
{compDeriv} & 528 & 205 & 323 (61.18\%)\\ \hline \hline
{\bf Total} & {\bf 40835} & {\bf 26632 } & {\bf 14203 } ({\bf 34.78\% }) \\ \hline\hline
\end{tabular}
\caption{The total number of student submissions and submissions with syntax errors for each problem.}
\tablabel{benchmarks}
\end{table}

\subsection{Training Phase}

During the training phase, we use all student submissions with no syntax errors for learning the token sequence model for a given problem. The student submissions are first tokenized into a set of sequence of tokens, which are then fed into the neural network for learning the token sequence model. The total number of tokens obtained from the syntactically correct student submissions for each problem in shown in \tabref{training}.  The table also presents the initial vocabulary size (the set of unique tokens in the student submissions) and the training vocabulary size, which is obtained by replacing all tokens whose occurrence frequency is under a threshold as \t{IDENT}. For our experiments, we use a threshold of $t=4$.

To train the recurrent neural network, we used a learning rate of $0.002$, a sequence length of $10$, and a batch size of $50$. We use the batch gradient descent method with rmsprop (decay rate 0.97) to learn the edge weights, where the gradients were clipped at a threshold of $5$. As we will see later, we experiment with both the RNN and LSTM networks with 1 or 2 hidden layers and each with either 128 or 256 hidden units. These neural networks were trained for a maximum of $40$ epochs and the time required to train each neural network for different problems was on an average $1.5$ hours. The experiments were performed on a 1.4 GHz Intel Core i5 CPU with 4GB RAM.

\begin{table}
\begin{tabular}{|c|c|c|c|c|}
\hline
{\bf Problem} & {\bf Correct} & {\bf Total} & {\bf Vocab} & {\bf Training}\\
& {\bf Attempts} & {\bf Tokens} & {\bf Size} & {\bf Vocab} \\ \hline
{recurPower} & 8176 & 338,958 & 191 & 117 \\ \hline
{iterPower} & 9194 & 358,849 & 795 & 526 \\ \hline
{oddTuples} & 8233 & 385,264 & 554 & 317 \\ \hline
{evalPoly} & 824 & 55,370 &  373 & 276 \\ \hline
{compDeriv} & 205 & 18,557 & 226 & 150 \\ \hline
\end{tabular}
\caption{The total number of tokens and the vocabulary size used for training the neural network.}
\tablabel{training}
\end{table}

\subsection{Number of Corrected Submissions}
We first present the overall results of our system in terms of how many student submissions are corrected using the predicted tokens in \tabref{overallresults}.
Since our algorithm currently considers only one syntax error in a student submission and there are many submissions with multiple syntax errors, we also report the number of cases where the suggested correction fixes the first syntax error but the submission isn't completely fixed because of other errors. We call this class of programs as \emph{Fixed(Other)}. In total, our system is able to provide suggestions to completely fix the syntax error in $31.69\%$ of the cases. Additionally, it is able to fix the first syntax error on a given error line without fixing other syntax errors on future lines in $6.39\%$ of the cases. The system isn't able to provide any fix to the errors for the remaining $61.92\%$ of the submissions. The number of programs that are completely and partially fixed for each individual problem is also shown in the table. 

We can first observe that even with relatively lesser number of total attempts for the \t{evalPoly} and \t{compDeriv} problems, the system is able to repair a significant number of syntax errors (40.43\%+11.73\% = 52.16\%). We do get some improvement with larger number of correct submissions, but the RNNs are able to learn comprehensive language models even with few hundreds of correct submissions. Another interesting observation is that the system is able to completely fix the syntax errors for a large fraction of the problems except for the \t{oddTuples} problem. On further manual inspection, we found that this was the case because the student attempts for the \t{oddTuples} problem consisted of a large number of indentation errors. Moreover, there was also a large number of diverse solution strategies that were not represented in the training set.

\begin{table}

\begin{tabular}{|c|c|c|c|c|}
\hline
{\bf Problem} & {\bf Incorrect} & {\bf Completely} & {\bf Fixed}\\
& {\bf Attempts} & {\bf Fixed} & {\bf (Other)}\\ \hline
{recurPower} & 2071 & 1061 (51.23\%) & 281 (13.57\%)\\ \hline
{iterPower} & 2661 & 1599 (60\%) & 276 (10.37\%) \\ \hline
{oddTuples} & 8824 & 1575 (17.85\%) & 303 (3.43\%) \\ \hline
{evalPoly} & 324 & 131 (40.43\%) & 38 (11.73\%) \\ \hline
{compDeriv} & 323 & 135 (41.79\%) & 10 (3.09\%) \\ \hline \hline
{\bf Total} & {\bf 14203 } & {\bf 4501 (31.69\%)} &{\bf 908 (6.39\%)} \\ \hline \hline
\end{tabular}

\caption{The number of submissions that are completely fixed and partially fixed (error in another line) by our system using the LSTM-(2,128) neural network.}
\tablabel{overallresults}
\end{table}

\begin{table*}
\centering
\begin{tabular}{|c|c|c|c|c|c|c|c|c|c|c|}
\hline
\multirow{3}{*}{\bf Problem} & \multirow{2}{*}{\bf Incorrect} & \multicolumn{5}{|c|}{\bf Completely Fixed} & \multicolumn{4}{|c|}{\bf Fixed (Other Line)}\\ \cline{3-11}
& &\multicolumn{2}{|c|}{\bf Offset} & \multicolumn{2}{|c|}{\bf Offset-1} & \multirow{2}{*}{\bf PrevLine} & \multicolumn{2}{|c|}{\bf Offset} & \multicolumn{2}{|c|}{\bf Offset-1}  \\ \cline{3-6} \cline{8-11} 
& {\bf Attempts} &{\bf Insert} & {\bf Replace} & {\bf Insert} & {\bf Replace} & & {\bf Insert} &{\bf Replace} & {\bf Insert} & {\bf Replace} \\ \hline
recurPower & 2071 & 48 & 48 & 467 & 708 & 856 & 16 & 15 & 215 & 310\\ \hline
iterPower & 2661 & 9 & 8 & 672 & 869 & 1206 & 191 & 214 & 241 & 360\\ \hline
oddTuples & 8824 & 29 & 28 & 464 & 622 & 1368 & 11 & 15 & 306 & 351\\ \hline
evalPoly & 324 & 7 & 5 & 44 & 47 & 108 & 15 & 15 & 30 & 40\\ \hline
compDeriv & 323 & 1 & 1 & 43 & 71 & 99 & 3 & 3 & 13 & 20 \\ \hline \hline
{\bf Total} & 14203 & 94 & 90 & 1690 & 2371 & 3637 & 236 & 262 & 805 & 1081 \\ \hline\hline
\end{tabular}
\caption{The breakdown of the method used for using the predicted token sequence to fix the syntax error in the original program. Offset corresponds to using the exact error location pointed out by the parser to construct the token prefix, Offset-1 corresponds to a token before the error token, Insert and Replace respectively denote whether the predicted token sequence is inserted or replaced, and PrevLine uses the prefix upto the previous line and replaces the error line with the predicted token sequence.}
\tablabel{correctionmethod}
\end{table*}

A more detailed breakdown of the number of submissions corrected or partially corrected by our system is shown in \tabref{correctionmethod}. The table reports the number of cases for which the syntax errors were fixed by the predicted token sequences using five different algorithmic choices: i) {\bf Offset:} the prefix token sequence is constructed from the start of the program to the error token reported by the parser, ii) {\bf Offset-1:} the prefix token sequence is constructed upto one token before the error token, iii) {\bf PrevLine}: the prefix token sequence is constructed upto the previous line and the error line is replaced by the predicted token sequence, (iv) {\bf Insert}: the predicted token sequence is inserted at the Offset location, and (v) {\bf Replace}: the original tokens starting at the Offset location are replaced by the predicted token sequence. As we can see, there is no one single choice that works better than every other choice. This motivates the need of the $\fixalgo$ algorithm that tries all these different algorithmic choices in the order of first finding an insertion or a replacement fix using the predicted token sequences of increasing length and then using the Previous Line method. We use this ranking order over the choices to prefer smaller changes to the original program.

We can observe that the Previous Line choice performs the best for the completely fixed case. The reason for this is that the algorithm has more freedom to make larger changes to the original program by completely replacing the error line. It also sometimes lead to undesirable semantic changes, which may not correspond to student's intuition. The Previous line changes are explored only after trying out the Insertion/Replace choices in the $\fixalgo$ algorithm. The replacement of original tokens with the predicted token sequences performs a little better than the insertion choice. Another interesting observation is that generating the prefix token sequences for querying the language model that end at one token earlier than the error token (Offset-1) performs a lot better than using prefix sequences that end at the error token (Offset). Finally, we observe that there are many student submissions that are fixed uniquely by each one of the 5 choices, and the algorithm therefore needs to consider all the choices.

There are about $28\%$ additional student submissions (amongst the $61.92\%$ of the submissions for which our technique can not generate any repair) for which we can provide some repair feedback by using the \t{PrevLine} choice. In these submissions, the replacement of the erroneous line with the predicted line causes the error to be fixed in the error line but does not necessarily make the program syntactically correct. We do not report these cases in the earlier tables as part of the partially fixed programs because often times the replaced line itself introduces new syntax errors in the submission. However, we believe that providing such fixes might still be beneficial to the students to provide them hints regarding the likely statements that should occur in place of the error line.

Another interesting point to note is that in some cases the number of partially fixed programs that are reported in \tabref{correctionmethod} is more than the number of partially fixed programs in \tabref{overallresults}. For instance for the \t{recurPower} problem, the Offset-1 and Replace combination can partially fix 310 submissions, whereas the number of partially fixed submissions reported in \tabref{overallresults} is 281 for the \t{recurPower} problem. This is the case because some of those $310$ submissions get completely corrected using other algorithmic choices and are instead counted in the Completely Fixed category in \tabref{overallresults}. 

\subsection{Different Neural Network Baselines}

In this section, we compare different baseline neural networks for learning the token sequence models and their respective effectiveness in correcting the syntax errors for the \t{recurPower} problem. In particular, we consider 6 baselines: (i) RNN-(1,128), (ii) RNN-(2,128), (iii) LSTM-(1,128), (iv) LSTM-(2,128), (v) LSTM-(1,256), and (vii) LSTM-(2,256), where the network NN-(x,y) denotes a neural network (RNN or LSTM) consisting of x number of hidden layers with y number of units each. The results for the 6 baseline networks is presented in \tabref{baselines}. There isn't a large difference amongst the performance of different neural networks. The RNN-(1,128) model fixes the largest number of student submissions completely ($1078$), and also has the best performance after including the partially corrected submissions. Interestingly, adding an additional hidden layer with more number of hidden units actually degrades the performance of the network on our dataset. Our hypothesis for this phenomenon is that the network with more hidden layers and more number of hidden units overfits the token sequences in the training phase and doesn't generalize as well as the neural network with fewer hidden units. Another interesting observation is that RNNs perform slightly better in our scenario of fixing syntax errors as compared to the LSTMs.

\begin{table}
\centering
\begin{tabular}{|c|c|c|c|c|}
\hline
{\bf Baseline} & {\bf Total} & {\bf Completely} & {\bf Fixed} & {\bf Total} \\ 
{\bf Network} & {\bf Incorrect} & {\bf Fixed} & {\bf (Other)} & {\bf Fixed} \\ \hline
RNN-(1,128) & 2071 & {\bf 1078} & 287 & {\bf 1365}\\ \hline
RNN-(2,128) & 2071 & 1062 & 267 & 1329\\ \hline
RNN-(2,256) & 2071&  990 & 302 & 1292\\ \hline
LSTM-(1,128) & 2071 & 1028 & 294 & 1322\\ \hline
LSTM-(2,128) & 2071 & 1061 & 281 & 1342\\ \hline
LSTM-(2,256) & 2071 & 1045 & 293 & 1338\\ \hline
\end{tabular}
\caption{The performance of different neural networks on the \t{recurPower} student submissions.}
\tablabel{baselines}
\end{table}

\subsection{Effect of Different Threshold values}

We also experiment with the performance of our system by varying the threshold values for constructing the training vocabulary. The results for 3 different threshold values ($t=1,4,8$) are shown in \tabref{threshold}. As the threshold increases, a larger number of tokens are now labeled as the \t{IDENT} token and thereby decreases the size of the training vocabulary. Without using any threshold value (t=1), the system fixes the fewest number of syntax errors for the \t{recurPower} problem. There are several incorrect submissions that cannot be corrected in this case because the learnt model performs poorly on prefix token sequences consisting of rarely used tokens (such as infrequent variable names). We can also observe that the threshold value of $4$ performs better than the threshold value of $8$. One hypothesis for this phenomenon is that the neural network over-generalizes some of the tokens to \t{IDENT} and loses the specific token information needed for fixing some syntax errors.

\begin{table}
\centering
\begin{tabular}{|c|c|c|c|c|}
\hline
{\bf Threshold} & {\bf Initial} & {\bf Training} & {\bf Completely} & {\bf Fixed} \\ 
{\bf Values} & {\bf Vocab} & {\bf Vocab} & {\bf Fixed} & {\bf (Other)} \\ \hline
t=1 & 191 & 191 & 904 & 307 \\ \hline
t=4 & 191 & 117 & 1078 & 287  \\ \hline
t=8 & 191 & 86 & 1068 & 277 \\ \hline
\end{tabular}
\caption{The performance of different threshold values for constructing the training vocabulary for the \t{recurPower} problem using the RNN-(1,128) neural network.}
\tablabel{threshold}
\end{table}
\section{Related Work}
In this section, we describe several related work on learning language models for Big Code, automated code grading approaches, and machine learning based grading techniques for other domains. 

{\bf Language Models for Big Code: } The most closely related work to ours is that on learning language models of source code from a large code corpus and then using these models for several applications such as learning natural coding conventions, code suggestions and auto-completion, improving code style, suggesting variable and method names etc. Hindle et al.~\cite{naturalness} use an n-gram model to capture the regularity of local project-specific contexts in software. They apply the learnt model to present suggestions for next tokens in the context of the Java language, and showed that the simple language model even without syntax or type information outperformed the state-of-the-art Eclipse IDE token suggestion engine. Nguyen et al.~\cite{nguyen} extended this syntactic n-gram language model to also include semantic token annotations that describe the token type and their semantic role (such as variable, function call etc.) and combine it with topic modeling to obtain n-gram topic model that also captures global technical concerns of the source files and pairwise association of code tokens. They apply this enhanced model for code suggestion and show that it improves the accuracy over syntactic n-gram approach by 18-68\%. Allamanis et al.~\cite{codemining} applied this technique of learning n-gram models on a much larger code corpus containing over 350 million lines of code, and showed that using a large corpus for training these n-gram model can significantly increase their predictive capabilities.

\textsc{Naturalize}~\cite{codeconventions} is a language-independent framework that uses the n-gram language model to learn the coding convention and coding style from a project codebase, and suggests revisions to improve stylistic consistency. It was used to suggest natural identifier names and formatting conventions, and achieved 94\% accuracy for suggesting identifier names as its top suggestion. The framework constructs an input snippet using the abstract syntax tree for the identifier for which the suggestions are needed and selects n-grams from this snippet containing the  identifier. The n-grams containing that identifier are also selected from a training corpus (source codes from project whose conventions are to be adopted). The learnt n-gram model is then used for scoring all the possible candidates selected from the training corpus to replace the identifier in the input codebase. Allamanis et al.~\cite{namesuggestion} recently presented a technique for suggesting method and class names from its body and methods respectively using a neural probabilistic language model. The input to these models is a sequence of finite length which is mapped to some lower dimension(D) vector (word representation). These vectors are then fed to hidden layer (non linear function) of the neural network. The final output layer is a softmax layer that evaluates the conditional probability of each word in the vocabulary given the input sequence. JSNice~\cite{bigcode} is a scalable prediction engine for predicting identifier names and type annotation of variables in JavaScript programs. The key idea in JSNice is to transform the input program into a representation that enables formulating the problem of inferring identifier names and type annotations as structured prediction with conditional random fields (CRFs). Given an input program, it first converts the program into a dependency network that captures relationships between program elements whose properties are to be predicted with elements whose properties are known. It then uses Maximum a Posteriori (MAP) inference to perform the structured prediction on the network.

Our technique is inspired from these previous work in learning language models from a corpus of code, but differs from them in four key ways. First, our application of using the language model to compute fixes to syntax errors in student submissions is different from the applications considered by previous approaches such as suggestions for identifier, method, and class names, code auto-completion and suggestion, and coding convention inference. Second, we use a recursive neural network (RNN) to capture long context relationships amongst tokens in a token sequence unlike previous approaches that use n-gram models, CRFs, and log bilinear neural networks. RNNs are traditionally considered to be hard to learn, but we leverage the recent advances in efficiently learning the RNNs for learning the token sequence models. Third, since we cannot obtain abstract syntax trees for programs with syntax errors, many of these techniques that depend on analyzing the ASTs are not applicable in our setting. Finally, we learn different RNN models for different programming assignments as opposed to learning a single model from the whole corpus. This allows us to find more accurate repairs for syntax errors that are problem dependent.

{\bf Automated Code Grading and Feedback:} The automated grading approaches can broadly be classified into two broad categories: 1) programming languages based approaches, and 2) machine learning based approaches. AutoProf~\cite{autoprof} is a system for providing automated feedback on functional correctness of introductory programming assignments. In addition to an incorrect student submission, it also takes as input a reference implementation specifying the intended functional behavior of the programming problem, and an error model consisting of rewrite rules corresponding to common mistakes that students make for the given problem. AutoProf uses constraint-based synthesis techniques~\cite{sketch} for finding minimum number of changes (guided by an error model) in the incorrect student submission to make it functionally equivalent to the reference implementation. Another approach~\cite{fse14} based on dynamic program analysis was recently presented for providing feedback on performance problems. It runs student submissions on a set of test cases to capture certain key values that occur during program executions, which are then used to identify the high-level strategy used by the student submission and provide corresponding feedback.

There has also been a lot of interest in the machine learning community on automated feedback generation and grading for programming problems. Huang et al.~\cite{moocshop} present an approach to automatically cluster syntactically similar Matlab/Octave programs based on the AST tree edit distance using an efficient approach based on dynamic programming. Codewebs~\cite{codewebs} creates a queryable index that allows for fast searches of code phrases into a massive dataset of student submissions. It accepts three form of queries: subtrees, subforests, and contexts that are subgraphs of an AST. A teacher provides detailed feedback on a few handful of labeled submissions, which is then propagated to thousands of student submissions by understanding the underlying structure present in the labeled submissions and querying the search engine.
Another recently proposed approach uses neural networks to simultaneously embed both the precondition and postcondition spaces of a set of programs into a feature space, where programs are considered as linear maps on this space. The elements of the embedding matrix of a program are then used as code features for automatically propagating instructor feedback at scale~\cite{icml15}.

The key difference between our technique and the previous programming languages and machine learning based techniques is that the previous techniques rely on the ability to generate the AST for the student submission to perform further analysis. However, with syntax errors, it is not possible to obtain such ASTs and that unfortunately limits these techniques to provide feedback on syntactic errors in student submissions.

{\bf Machine learning for Grading in Other domains:} There have been similar automated grading systems developed for domains other than programming such as Mathematics and short answer questions. The Mathematical Language Processing (MLP)~\cite{mathfeedback} framework leverages solutions from large number of learners to evaluate correctness of student solutions to open response Mathematical questions. It first converts an open response to a set of numerical features, which are then used for clustering to uncover structures of correct, partially-correct, and incorrect solutions. A teacher assigns a grade/feedback to one solution in a cluster, which is then propagated to all solutions in the cluster. Basu et al.~\cite{powergrading} present an approach to train a similarity metric between short answer responses to United States Citizenship Exam, which is then used to group the responses into clusters and subclusters for powergrading. The main difference between our technique and these techniques is that we use RNNs to learn a language model for token sequences unlike machine learning based clustering approaches used by these techniques. Moreover, we focus on giving feedback on syntax errors whereas these techniques focus on semantic correctness of the student solutions.

\section{Limitations and Future Work}

There are several limitations in the presented algorithm that we would like to extend in future work. One limitation of our technique is that it currently handles only one syntax error in the student program. For example, consider the student submission in \figref{multipleerrors}. The $\fixalgo$ algorithm is able to correctly fix the first indentation error in line 3 by inserting a tab token before the return token, but the updated program does not pass the compiler check because of another indentation error in line 5. We plan to extend our algorithm to also handle multiple syntax errors by automating the process of fixing the first syntax error found in the program using the $\fixalgo$ algorithm and then calling the algorithm again recursively on the next error found in the updated program.

\begin{figure}
\begin{python}
def recurPower(base, exp):
    if exp == 0:
    return 1
    else:
    return base * base**(exp-1)
\end{python}
\caption{A student submission with multiple syntax errors in lines 3 and 5.}
\figlabel{multipleerrors}
\end{figure}

Our system currently only uses the prefix token sequence for suggesting the token sequence for repair. For the program shown in \figref{bidirectional}, the algorithm suggests the fix corresponding to the expression \t{exp==0}. If the algorithm also took into account the token sequences following the error location such as \t{return base}, then it could have predicted a better fix corresponding to the token sequence \t{exp == 1}. There is a class of RNN called bidirectional-RNN that allows for predicting tokens based on both past and future contexts. We intend to investigate in future work if bidrectional-RNNs can be trained efficiently in our setting and if they can improve the fix coverage.

\begin{figure}
\begin{python}
def recurPower(base, exp):
    if exp = 0:
        return base
    return base * recurPower(base, exp-1)
\end{python}
\caption{ The fix generated by our system suggests the expression \t{exp == 0} using the prefix token sequence. A better fix \t{exp == 1} can be suggested if the algorithm also took into account the token sequence following the error location \t{return base}.}
\figlabel{bidirectional}
\end{figure}

Another limitation of our technique is that it only checks for syntactic correctness while finding a repair candidate. There are some cases where the suggested sequence fixes the syntax errors but is semantically incorrect. We can try to solve this issue by adding a semantic check in the $\fixalgo$ algorithm in addition to the syntactic parser check, and by allowing the algorithm to query the learnt model for multiple token sequence predictions until we obtain one that is a semantically correct fix as well.

Finally, there is an interesting research question on how to best translate the repairs generated by our technique into good pedagogical feedback, especially the cases for which the suggested fix is not semantically correct. Some syntax errors are simply typos such as mismatched parenthesis or missing operators, for which the feedback generated by our technique should be sufficient. But there are some class of syntax errors that point to deeper misconceptions in the student's mind. Some examples of such errors include assigning to return keyword e.g. \t{return = exp}, performing an assignment inside a parameter value of a function call e.g. \t{recurPower(base,exp-=1)}, etc. We would like to build a system on top of our technique that can first distinguish small syntax errors from deeper misconception errors, and then translate the suggested repair fix accordingly so that students can learn the high-level concepts for correctly understanding the language syntax.
\section{Conclusion}

In this paper, we presented a technique to use Recurrent neural networks (RNNs) to learn token sequence models for finding repairs to syntax errors in student programs. For a programming assignment, our technique first uses the set of all syntactically correct student submissions to train an RNN for learning the token sequence model, and then uses the trained model to predict token sequences for finding repairs for student submissions with syntax errors. Our technique takes inspiration from two emerging research areas: 1) Learning language models from big code, and 2) Efficient learning techniques for Recurrent neural networks. For our dataset of student attempts obtained from the edX platform, our technique can generate repairs for about 32\% of submissions. We believe this technique can provide a basis for providing automated feedback on syntax errors to hundreds of thousands of students learning from online introductory programming courses that are being taught by edX, Coursera, and Udacity.

\bibliography{references}
\bibliographystyle{abbrv}

\end{document}